# Adaptive-Aggressive Traders Don't Dominate

Daniel Snashall[0000-0003-3051-7716] and Dave Cliff[0000-0003-3822-9364]

Department of Computer Science, University of Bristol, Bristol BS8 1UB, U.K.

ds15012.2015@my.bristol.ac.uk and csdtc@bristol.ac.uk

**Abstract.** For more than a decade Vytelingum's *Adaptive-Aggressive* (AA) algorithm has been recognized as the best-performing automated auction-market trading-agent strategy currently known in the AI/Agents literature; in this paper, we demonstrate that it is in fact routinely outperformed by another algorithm when exhaustively tested across a sufficiently wide range of market scenarios. The novel step taken here is to use large-scale compute facilities to brute-force exhaustively evaluate AA in a variety of market environments based on those used for testing it in the original publications. Our results show that even in these simple environments AA is consistently outperformed by IBM's *GDX* algorithm, first published in 2002. We summarize here results from more than one million market simulation experiments, orders of magnitude more testing than was reported in the original publications that first introduced AA. A 2019 ICAART paper by Cliff claimed that AA's failings were revealed by testing it in more realistic experiments, with conditions closer to those found in real financial markets, but here we demonstrate that even in the simple experiment conditions that were used in the original AA papers, exhaustive testing shows AA to be outperformed by GDX. We close this paper with a discussion of the methodological implications of our work: any results from previous papers where any one trading algorithm is claimed to be superior to others on the basis of only a few thousand trials are probably best treated with some suspicion now. The rise of cloud computing means that the compute-power necessary to subject trading algorithms to millions of trials over a wide range of conditions is readily available at reasonable cost: we should make use of this; exhaustive testing such as is shown here should be the norm in future evaluations and comparisons of new trading algorithms.

**Keywords:** Automated Trading, Auction Markets, Adaptive Bidding Agents.

## 1    Introduction

For hundreds of years regional, national, and international financial markets involved human traders interacting with one another to negotiate and agree details of transactions. In the past 15 years the number of human traders in financial markets has fallen very sharply, as humans have been systematically replaced by automated trading systems. These automated systems, known in the industry as "algorithmic traders" (often abbreviated simply to "algos") or "robot traders", can employ artificial intelligence (AI) and machine learning (ML) techniques to adapt their responses over multiple timescales ranging from milliseconds to years. In a large investment bank, a single robot trader might routinely handle daily order flows of US$20Bn or more. This is manifestly a big business, and the replacement of highly-paid humans (whom, it seems reasonable to assume, were also highly intelligent) with more cost-efficient robot traders is potentially a notable success story for AI/ML. Major investment banks and fund-management companies no longer compete to hire only the best traders; now they compete to hire the best trading-algorithm designers too. See [16] for an entertaining first-hand account of these changes.

Because of the large sums of money at stake, precise details of the specific robot traders used in industry are closely guarded commercial secrets. If a robot is making millions of dollars for a bank, the last thing the bank wants is for someone to publish an academic paper describing how that robot works: any commercial advantage would be immediately lost. Nevertheless, there is a body of work in the academic AI/ML literature stretching back to the late 1990s that describes a sequence of adaptive automated trading algorithms which have stood the test of time and remain influential to this day.

Although a few significant publications contributing to the development of robot-trading systems came from academic economists, the landmark papers largely appeared in AI and autonomous-agent publication venues such as the *International Joint Conference on Artificial Intelligence* (IJCAI), the *International Conference on Autonomous Agents and Multi-Agent Systems* (AAMAS), the *International Conference on Agents and Artificial Intelligence* (ICAART), and the prestigious *Artificial Intelligence* journal (AIJ): Section 2 reviews in more detail eight major publications in the development of this field. The review in Section 2 is important, because there we trace the way in which the methodology of initial experiments published in 1962 by a young economist, Vernon Smith (who 40 years later would be awarded the Nobel Prize for his empirical research work) have since come to be fixed, or fixated upon, in the AI/agents literature on robot traders. Motivated by what it seems fair to assume was a wholly well-intentioned desire to show each set of the latest results in the context of what had gone before, papers subsequent to Smith's replicated much or all of his 1962 experiment design and analysis. And this, it seems, may have led down something of a dead end.

More details are given in Section 2 but for the purposes of this introduction it is sufficient to summarize the key events as follows: at the 2001 IJCAI a team of researchers at IBM published results [7] which showed that two robot trading algorithms, known as *MGD* [23] and *ZIP* [3], could consistently out-perform human traders when tested in rigorous laboratory-style experiments; in the years after this, several other trading algorithms were published, each being claimed as the best-performing algorithm in the public domain at the time of its publication; and the most recent of these is Vytelingum's *AA* algorithm [26] which was described in a 2006 paper in the AIJ [27], and was later shown to outperform human traders in a 2011 IJCAI paper [9]. Put simply, AA is widely believed to be the best-performing trading algorithm in the published literature.

In this paper we demonstrate that belief to be wrong: we show here that AA is not the best. Our demonstration builds on recently-published work by Vach [25] and by Cliff [6]. As far as we are aware, Vach's 2015 MSc thesis [25] was the first to publicly question whether AA is indeed dominant: Vach reported results in which he populated markets with a variety of different robot traders (i.e., some traders running AA and other traders running different strategies, such as MGD or ZIP), that then interacted with one another; Vach found that whether AA was the best-performing algorithm or not in any particular trial depended on the relative proportions of the different trading agents present in the market for that trial. But if AA was truly dominant then it should have outperformed other robot traders regardless of what the mix of strategies is in the market at any one time. Inspired by Vach, and seeking to independently replicate his results, Cliff's 2019 ICAART paper [6] presented results from exhaustive brute-force testing in which, for a market with $N$ traders active in it, and with a selection of $T$ robot-trader algorithms (including AA) available, the performance of AA in every possible permutation of the $T$ different trader types was studied over a variety of values of $N$. Cliff's results, gathered from more than 3 million individual market simulation trials, confirmed and extended Vach's observation: for each value of $N$ that Cliff studied, there was some permutation of the $T$ different robot-trader strategies in which AA is outperformed by one or more of the other strategies. Cliff assumed that this result was attributable to his use of test environments that were more realistic (i.e., closer to real-world financial markets) than those that had been used by Vytelingum in his 2006 [26] and 2008 [27] publications introducing AA. In this paper we present results demonstrating that Cliff's assumption in [6] was incorrect. Here we go back to the original test-cases used by Vytelingum [26,27], but we follow Cliff's [6] method of running brute-force exhaustive testing of all possible permutations of AA and other strategies: whereas Vytelingum published results from fewer than 30,000 simulation trials, in this paper we show results from more than 1,000,000 market sessions: a 30-fold increase over the original publications. Our results here are consistent with those reported by Vach [25] and by Cliff [6]: AA can be routinely outperformed by other strategies, depending on the relative proportions of the different strategies in the market; thus the claims of AA's dominance in earlier publications seem now to be due entirely to an insufficient number of trials having been conducted, even in the original test-cases used in the initial publications on AA. If the exhaustive testing we used here had been conducted at the time of the original publications, AA would not have been mistakenly described as the best-known strategy.

The testing we use is not complicated: it just requires some nested loops to iterate through all possible permutations of the various trader-types, but its combinatorics are truly explosive and hence performing all the necessary trials is highly computationally expensive, and would have taken an awful long time on a single desktop computer. Possibly these high computational costs are why such exhaustive testing has not previously been commonplace in the evaluation of trading algorithms. For computing the brute-force simulation studies described here we used our University's in-house *Blue Crystal* supercomputer, to which we have free access; but all of our experiments could just as easily have been run instead on commercial cloud computing services such as those available from Amazon, Google, Microsoft, or Oracle, incurring only modest fees (a few hundred dollars at most, at today's prices). And so, while the results from our experiments constitute the *empirical* contribution in this paper, we also offer the style of testing used here as a *methodological* contribution: given the present-day ready availability of cheap large-scale computing via cloud service providers, we argue later in this paper that the kind of brute-force studies reported here should from now on be adopted as the norm in any work that evaluates and compares trading algorithms.

The rest of this paper is structured as follows. Section 2 covers the necessary background material, and Section 3 describes how AA can be modified to work in contemporary market simulators. The text in those two sections is taken verbatim from [6], and readers familiar with that paper can safely skip straight to Section 4, which is where we describe our methods and results for exhaustive testing of AA. Section 5 then discusses methodological implications, and conclusions are drawn in Section 6.

## 2 Traders, Markets, and Eight Key Papers

The 2002 Nobel Prize in Economics was awarded to Vernon Smith, in recognition of Smith's work in establishing and thereafter growing the field of *Experimental Economics* (abbreviated hereafter to "ExpEcon"). Smith showed that the microeconomic behavior of human traders interacting within the rules of some specified market, known technically as an *auction mechanism*, could be studied empirically, under controlled and repeatable laboratory conditions, rather than in the noisy messy confusing circumstances of real-world markets. The minimal laboratory studies could act as useful proxies for studying real-world markets of any type, but one particular auction mechanism has received the majority of attention: the *Continuous Double Auction* (CDA), in which any buyer can announce a bid-price at any time and any seller can announce an offer-price at any time, and in which at any time any trader in the market can accept an offer or bid from a counterparty, and thereby engage in a transaction. The CDA is the basis of most major financial markets worldwide.

Smith's initial set of experiments were run in the late 1950's, and the results and associated discussion were presented in his first paper on ExpEcon, published in the highly prestigious *Journal of Political Economy* (JPE) in 1962 [18]. It seems plausible to speculate that when his JPE paper was published, Smith had no idea that it would mark the start of a line of research that would eventually result in him being appointed as a Nobel laureate. And it seems even less likely that he would have foreseen the extent to which the experimental methods laid out in that 1962 paper would subsequently come to dominate the methodology of researchers working to build adaptive autonomous trading agents by combining tools and techniques from AI, ML, agent-based modelling (ABM), and agent-based computational economics (ACE). Although not a goal stated at the outset, this strand of AI/ML/ABM/ACE research converged toward a common aim: specifying an artificial agent, an autonomous adaptive trading strategy, that could automatically tune its behavior to different market environments, and that could reliably beat all other known automated trading strategies, thereby taking the crown of being the current best trading strategy known in the public domain, i.e., the "dominant strategy". Over the past 20 years the dominant strategy crown has passed from one algorithm to another. Here, we demonstrate that the current holder of the title, Vytelingum's [26, 27] *AA* strategy, does not perform nearly so well as was previously believed from earlier successes in small numbers of trials.

Given that humans who are reliably good at trading are generally thought of as being "intelligent" in some reasonable sense of the word, the aim to develop ever more sophisticated artificial trading systems is clearly within the scope of AI research, although some very important early ideas came from the economics literature: a comprehensive review of relevant early research was given in [3]. Below in Section 2.1 we first briefly introduce eight key publications leading to the development of AA; then describe key aspects of ExpEcon market models in Section 2.2; and then discuss each of the eight key publications in more detail in Section 2.3. After that, Section 2.4 summarizes the results of Vach [25] and Cliff [6], which together cast doubts on the hitherto apparently resolved issue of which trading agent is the best.

## 2.1 A Brief History of Trading Agents

If our story starts with Smith's 1962 JPE paper, then the next major step came 30 years later, with a surprising result published in the JPE by Gode and Sunder in 1993 [14]: this popularized a minimally simple automated trading algorithm now commonly referred to as *ZIC*. A few years later two closely related research papers were published independently and at roughly the same time, each written without knowledge of the other: the first was a Hewlett-Packard Labs technical report [3] describing the adaptive AI/ML trading-agent strategy known as the *ZIP* algorithm; the second summarized the PhD thesis work of Gjerstad, in a paper [11] co-authored with his PhD advisor Dickhaut, describing an adaptive trading algorithm now widely known simply as *GD*. After graduating his PhD, Gjerstad worked at IBM's TJ Watson Labs where he helped set up an ExpEcon laboratory that his IBM colleagues used in a study that generated worldwide media coverage when the results were published by Das *et al.* at IJCAI-2001 [7]. This paper presented results from studies exploring the behavior of human traders interacting with GD and ZIP robot traders, in a CDA with a Limit Order Book (LOB: explained in more detail in Section 2.2, below), and demonstrated that both GD and ZIP reliably outperformed human traders. Neither GD nor ZIP had been designed to work with the LOB, so the IBM team modified both strategies for their study. A follow-on 2001 paper [23] by Tesauro and Das (two co-authors of [7]) described a more extensively *Modified GD* (MGD) strategy, and later Tesauro and Bredin [23] described the *GD eXtended* (GDX) strategy. Both MGD and GDX were each claimed to be the strongest-known public-domain trading strategies at the times of their publication.

Subsequently, Vytelingum's 2006 thesis [26] introduced the *Adaptive Aggressive* (AA) strategy which, in an AIJ paper [27], and in later *ICAART* and *IJCAI* papers [8, 9], was shown to be dominant over ZIP, GDX, and also human traders. Thus far then, AA holds the title.

However Vach [25] presented results from experiments with the *OpEx* market simulator [10], in which AA, GDX, and ZIP were set to compete against one another, and in which the dominance of AA is questioned: Vach's results indicate that whether AA dominates or not can be dependent on the ratio of AA:GDX:ZIP in the experiment: for some ratios, Vach found AA to dominate; for other ratios, it was GDX. Vach studied only a relatively small sample from the space of possible ratios, but his results prompted Cliff [6] to exhaustively step through a wide range of differing ratios of four trading strategies (AA, ZIC, ZIP, and the minimally simple SHVR strategy described in Section 2.2), doing a brute-force search for situations in which AA is outperformed by the other strategies. The combinatorics of such a search are quite explosive: Cliff reported on results from over 3.4 million individual simulations of market sessions. Cliff's findings indicated that Vach's observation was correct: AA's dominance does indeed depend on how many other AA traders are in the market; and, in aggregate, AA was routinely outperformed by ZIP and by SHVR.

## 2.2 On Laboratory Models of Markets

Smith's early experiments were laboratory models of so called *open-outcry trading pits*, a common sight in any real financial exchange before the arrival of electronic trader-terminals in the 1970s. In a trading pit, human traders huddle together and shout out their bids and offers, and also announce their willingness to accept a counterparty's most recent shout. It was a chaotic scene, now largely consigned to the history books. In the closing quarter of the 20th Century, traders moved *en masse* to interacting with each other instead via electronic means: traders "shouted" their quote-prices (offer or bid) or acceptances by typing orders on keyboards and then sending those orders to a central server that would display an aggregate summary of all orders currently "shouted" (i.e., quoted) onto the market. That aggregate summary is very often in the form of a *Limit Order Book* or LOB: the LOB summarizes all bids and offers currently live in the market. At its simplest, the LOB is a table of numbers, divided into the *bid side* and the *ask side* (also known as the *offer side*). Both sides of the LOB show the best price at the top, with less good prices arranged below in numeric order of price: for the bid side this means the highest-priced bid at the top with the remaining bid prices displayed in descending order below; and for the ask side the lowest-priced offer is at the top, with the remaining offers arranged in ascending order below. The arithmetic mean of the best bid and best ask prices is known as the *mid-price*, and their difference is the *spread.* For each side of the LOB, at each price on the LOB, the quantity available on that side at that price is also indicated, but with no indication of who the relevant orders came from: in this sense the LOB serves not only to aggregate all currently live orders, but also to anonymize them.

Traders in LOB-based markets can usually cancel existing orders to delete them from the LOB. In a common simple implementation of a LOB, traders can accept the current best bid or best offer by issuing a quote that *crosses the spread*: i.e., by issuing an order that, if added to the LOB, would result in the best bid being at a higher price than the best ask. Rather than be added to the LOB, if a bid order crosses the spread then it is matched with the best offer on the ask side (known as *lifting the ask*), whereas an ask that crosses the spread is matched with the best bid (*hitting the bid*); and in either case a transaction then occurs between the trader that had posted the best price on the relevant side of the LOB, and the trader that crossed the spread. The price of the resulting transaction is whatever price was hit or lifted from the top of the LOB.

Smith's earliest experiments pre-dated the arrival of electronic trading in real financial markets, and so they can be thought of as laboratory models of open-outcry trading pits. Even though the much later work by Gode and Sunder [14], Cliff [3], Gjerstad and Dickhaut [11], and Vytelingum [26] all came long after the introduction of electronic LOBs in real markets, these academic studies all stuck with Smith's original methodology, of modelling open-outcry markets (often by essentially operating a LOB with the depth fixed at 1, so the *only* information available to traders is the current best, or most recent, bid and ask prices).

Nevertheless, the studies by IBM researchers [7, 23, 24], and also the replication and confirmation of AA results by De Luca and Cliff [8-10] and by Stotter *et al.* [21, 22], all used LOB-based market simulators. The IBM simulator *Magenta* seems to have been proprietary to IBM; developed at TJ Watson Labs and not available for third-party use, but De Luca made an open-source release of his *OpEx* simulator [10] which was subsequently used by Vach [25] in the studies that prompted our work reported here.

Also of relevance here is the *ExPo* simulator described by Stotter *et al.* [21, 22]: in the work by De Luca [8-10], by Vach [25], and by Stotter *et al.* [21, 22], Vytelingum's original AA needed modifications to make it work in a LOB-based market environment: this is discussed further in Section 3.

In the work reported here we used neither OpEx nor ExPo, but instead *BSE* [1, 5] which is another open-source ExpEcon market simulator, initially developed as a teaching aid but subsequently used as a platform for research (see, e.g. [15]). BSE has the advantage of being relatively lightweight (a single Python script of c.2500 lines) and hence readily deployable over large numbers of virtual machines in the cloud. BSE maintains a dynamically updated LOB and also publishes a *tape*, a time-ordered record of all orders that have been executed, and other significant events such as the cancellation of earlier orders (which are deleted from the LOB). BSE comes with pre-defined versions of ZIC and ZIP, and also some additionally minimally-simple non-adaptive trading strategies that can be used for benchmarking against other more complex strategies added by the user. One of these, the *Shaver* strategy (referred to in BSE by the "ticker symbol" SHVR) simply reads the best prices on the LOB and, if it is able to do so without risking a loss-making deal, then issues an order that improves the current best bid or best ask by 0.01 units of currency (i.e., one penny/cent), which is BSE's *tick size*, i.e. the minimum change in price that the system allows.

## 2.3 Eight Key Papers, One Methodology

**Smith, 1962.** Although precedents can be pointed to, Smith's 1962 JPE paper [18] is widely regarded as the seminal study in ExpEcon. In it he reported on experiments in which a group of c.12-25 human subjects were each randomly assigned to be either a *buyer* or a *seller* in the market experiment. Buyers were given a supply of artificial money, and sellers were given one or more identical items, of no intrinsic value, to sell. Each trader in the market was assigned a private valuation, a secret *limit price*: for a buyer this was the price above which he or she should not pay when purchasing an item; for a seller this was the price below which he or she should not sell an item. These limit-price assignments model the client orders executed by sales traders in real financial markets; we'll refer to them just as *assignments* in the rest of this paper. After the allocation of assignments to all traders, the traders then interacted via an open-outcry CDA while Smith and his assistants made notes on the sequence of events that unfolded during the experiment: typically, buyers would gradually increase their bid-prices, and sellers would gradually lower their offer-prices (also known as ask-prices) until transactions started to occur. Eventually, usually within a few minutes, the experimental market reached a position in which no more trades could take place, which marked the end of a *trading period* or "trading day" in the experiment; any one experiment typically ran for *n*=5-10 periods, with all the traders being resupplied with fresh assignments of limit prices and money-to-buy-with and items-for-sale at the start of each trading period. The sequence of *n* contiguous trading periods (or an equivalently long single-period experiment with continuous replenishment, as discussed further in Sections 2.4 and 4.4) is referred to here as one *market session*. Smith could induce specific supply and demand curves in these experimental markets by appropriate choices of the various limit-prices he assigned to the traders. As any high-school student of microeconomics knows, the market's theoretical *equilibrium price* (denoted hereafter by $P_0$) is given by the point where the supply curve and the demand curve intersect. Smith found that, in these laboratory CDA markets populated with only remarkably small groups of human traders, transaction prices could reliably and rapidly converge on the theoretical $P_0$ value despite the fact that each human trader was acting purely out of self-interest and knew only the limit price that he or she had been assigned. Smith's analysis of his results focused on a statistic that he referred to as $\alpha$, the root mean square deviation of actual transaction prices from the $P_0$ value over the course of an experiment. In his early experiments, $P_0$ was fixed for the duration of any one experiment; in later work Smith explored the ability of the market to respond to "price shocks" where, in an experiment of $N$ trading days, on a specific day $S<N$ the allocation of limit prices would be changed, altering $P_0$ from the value that had been in place over trading periods *1, 2, ..., S*, to a different value of $P_0$ that would then remain constant for the rest of the experiment, i.e. in trading periods *S+1, S+2, ..., N*. For brevity, in the rest of this paper Smith's initial style of experiments will be referred to as *S'62* experiments.

**ZIC: Gode and Sunder, 1993.** Gode and Sunder's JPE paper [14] used the S'62 methodology, albeit with the CDA markets being electronic (a move Smith himself had made in his experiments many years earlier), so each trader was sat at a personal terminal, a computer screen and keyboard, from which they received all information about the market and via which they announced their orders, their bids or offers, to the rest of the traders in the experiment. Gode and Sunder first conducted a set of experiments in which all the traders were human, to establish baseline statistics. Then, all the human traders were replaced with automated trading systems, absolute-zero minimally-simple algo traders which Gode and Sunder referred to as *Zero Intelligence* (ZI) traders. Gode and Sunder studied markets populated with two type of ZI trader: *ZI-Unconstrained* (ZIU), which simply generated random prices for their bids or offers, regardless of whether those prices would lead to profitable transactions or to losses; and *ZI-Constrained* (ZIC), which also generated random order prices but were constrained by their private limit prices to never announce prices that would lead them to loss-making deals. Gode and Sunder used fixed supply and demand schedules in each experiment, i.e. there were no price-shocks in their experiments.

Not surprisingly, the market dynamics of ZIU traders were nothing more than noise. But the surprising result in Gode and Sunder's paper was the revelation that a commonly used metric of market price dynamics known as *allocative efficiency* (AE, hereafter) was essentially indistinguishable between the human markets and the ZIC markets. Because AE had previously been seen as a marker of the degree to which the traders in a market were behaving intelligently, the fact that ZIC traders scored AE values largely the same as humans was a shock. Gode and Sunder proposed that a different metric should instead be used as a marker of the intelligence of traders in the market. This metric was *profit dispersion* (PD, hereafter) which measures the difference between the profit each trader accrued in an experiment, compared to the profit that would be expected for that trader if every transaction in the market had taken place at the market's theoretical equilibrium price $P_0$: humans typically showed very low values of PD (which is assumed to be good) while ZIC traders did not. On this basis, Gode and Sunder argued that PD should be used in preference to AE in future.

Other researchers were quick to cite Gode and Sunder's ZIC result, and often used it to support the claim that, given the ZIC traders have no intelligence, then for transaction prices to converge toward the theoretical equilibrium price and/or for a group of traders to score highly on AE, somehow the "intelligence" required to do this must reside within the rules of the CDA market system rather than within the heads of the traders. Strangely, Gode & Sunder's 1993 paper [14] provides no concrete causal mechanistic explanation of how their striking ZIC results arise; they describe their methods, and the results observed, but the internal mechanisms that give rise to those results are left as something of a mystery, as if the CDA market was an impenetrable black-box.

A causal mechanistic analysis of markets populated by ZIC traders was subsequently developed by Cliff [3], who considered the probability mass functions (PMFs) of prices generated by ZIC buyers and sellers, and the joint PMF of transaction prices in ZIP markets, which is given by the intersection of the bid-price and offer-price PMFs: the shape of the transaction-price PMF is determined by the nature of the supply and demand curves in the market, and Cliff demonstrated that the supply and demand curves in a ZIC market experiment could be arranged so that the expected value of the transaction prices (computable as an integral over the PMF) is identical to the theoretical equilibrium price given by the intersection point of the supply and demand curves. This was why the five ZIC experiments reported in Gode and Sunder's [14] paper showed transaction prices that were centered on the theoretical equilibrium price in each case: the supply and demand curves were arranged in such a way that this was the expected outcome. Cliff showed that with different arrangements of supply and demand curves, such as situations where one or both curves were flat (as had been used in Smith's original 1962 JPE paper [18]), the expected price of transactions in ZIP markets could differ considerably from the theoretical equilibrium price, and so transaction prices in those ZIC markets would fail to exhibit human-like convergence toward the theoretical equilibrium value. In these differently-designed experiments, ZIC traders would be revealed for exactly what they are: simple stochastic processes that only coincidentally exhibit human-like market dynamics when the experimenters happen to have chosen to impose just the right kind of supply and demand curves. Cliff's analysis showed that the level of intelligence in the ZIC traders was insufficient to recreate human-like market dynamics more broadly, and so a more intelligent automated trading strategy was required.[1]

**ZIP: Cliff, 1997.** Taking direct inspiration both from Smith's work and from the ZI paper by Gode and Sunder, Cliff [3] developed a ZI trading strategy that used simple machine-learning techniques to continuously adapt the randomly-generated prices quoted by the traders: this strategy, known as ZI-Plus (ZIP) was demonstrated to show human-like market dynamics in experiments with flat supply and/or demand curves: Cliff also showed theoretical analyses and empirical results which demonstrated that transaction prices in markets populated only by ZIC traders would not converge to the theoretical equilibrium price when the supply and/or demand curves are flat (or, in the language of microeconomics, "perfectly elastic"). ExpEcon studies in which the supply and/or demand curve was flat had previously been reported by Smith and others, but Gode and Sunder had not explored the response of their ZIC traders to this style of market. Cliff's work involved no human traders: all the focus was on markets populated entirely by autonomous agents, by ZIP traders. In total Cliff [3] reported on fewer than 1,000 simulated market sessions. The focus on *homogenous* markets can fairly be interpreted as continuing the tradition established by Gode and Sunder (who studied markets homogeneously populated with either human, ZIU, or ZIC traders) and by Smith (who studied all-human markets). In all other regards Cliff continued the S'62 tradition: key metrics were Smith's $\alpha$, AE, and PD.

---

[1] Independently, and via a wholly different line of attack, Gjerstad and Shachat [13] also demolished the argument that Gode and Sunder's [14] ZIC results indicate that the efficiency or intelligence in the market system lies solely within the CDA mechanism. Nevertheless, Gode and Sunder's results continue to be cited uncritically by various authors in the economics literature: we can only assume that such authors prefer a nice fairy story, rather than hard facts.

**GD: Gjerstad and Dickhaut, 1997.** Gjerstad's PhD studies of price formation in CDA markets also involved creating an algorithm that could trade profitably by adapting its behavior over time, in response to market events [11]. In contrast to the ZI work, Gjerstad's trading algorithm uses frequentist statistics, gradually constructing and refining a *belief function* that estimates the likelihood for a bid or offer to be accepted in the market at any particular time, mapping from price of the order to its probability of success. Gjerstad did not explicitly name his strategy, but it has since become known as the GD strategy. In all other regards, as with Cliff's work [3] and Gode and Sunder's [14], Gjerstad's [11] work was firmly in the S'62 tradition: homogenous markets of GD traders interacting in a CDA, buying and selling single items, with the metrics being Smith's $\alpha$, AE, and PD. In a later paper [12], Gjerstad made some refinements to the GD algorithm, adding a time-sensitivity or *pace* parameter, and named it *HBL* (for Heuristic Belief Learning), although the original GD form remains by far the most cited.

**MGD: Das *et al.*, 2001.** In their landmark 2001 IJCAI paper [7], IBM researchers Das, Hanson, Kephart, and Tesauro studied the performance of GD and ZIP in a series of ExpEcon market experiments where, for the first time ever in the same market, some of the traders were robots while others were human (recall that the earlier work of Smith, of Gode and Sunder, of Cliff, and of Gjerstad and Dickhaut had all studied homogeneous markets: either all-human or all-robot). Das *et al.* used a LOB-based market simulator called *Magenta*, developed by Gjerstad, and ran a total of six experiments, six market sessions, in which humans and robots interacted and where there were three shock-changes to $P_0$, i.e. four phases in any one experiment, each phase with a different $P_0$ value that was held static over that phase. The surprising result in this paper was that robot trading strategies could consistently outperform human traders, by significant margins: a result that attracted worldwide media attention. Both GD and ZIP outperformed human traders, and in the six experiments reported by Das *et al.* the results from the two robot strategies are so similar as to not obviously be statistically significant. A subsequent paper by IBM's Tesauro and Das [23], reported on additional studies in which a *Modified GD* (MGD) strategy was exhibited what the authors described in the abstract of their paper as *"...the strongest known performance of any published bidding strategy"*.

**GDX: Tesauro and Bredin, 2002.** Extensions to MGD were reported by IBM researchers Tesauro and Bredin at AAMAS 2002 [24]. This paper described extensions to MGD, using dynamic programming methods: the extended version was named *GDX* and its performance was evaluated when competing in heterogenous markets with ZIP and other strategies. Tesauro and Bredin reported that GDX outperformed the other strategies and claimed in the abstract of their paper that GDX *"…may offer the best performance of any published CDA bidding strategy.*"

**AA: Vytelingum, 2006.** Vytelingum developed AA and documented it in full in his PhD thesis [26] and in a major paper in the *AIJ* [27]. The internal mechanisms of AA are described in greater detail in Section 3 of this paper. Although Vytelingum's work came a few years after the IBM publications, the discussion within Vytelingum's publications is phrased very much in terms of the S'62 methodology: the $P_0$ value in his AA experiments was either fixed for the duration of each market session, or was subjected to a single "price shock" partway through the session (as described in Section 2.3.1); and again the primary metrics studied are Smith's $\alpha$, AE, and PD. Vytelingum presented results from heterogeneous market experiments where AA, GDX, and ZIP traders were in competition, and the published results indicated that AA outperformed both GDX and ZIP by small margins. In total, results from c.25,000 market sessions are presented in [27].

**AA Dominates: De Luca and Cliff, 2011.** As part of the research leading to his 2015 PhD thesis [10], De Luca used his LOB-based *OpEx* market simulator system to study the performance of AA in heterogeneous market experiments where some of the traders were AA, some were other robot strategies such as ZIP, and some were human traders sat at terminals interacting with the other traders (human and robot) in the market via the OpEx GUI, in the style introduced by the IBM team in their IJCAI 2001 paper. De Luca and Cliff [8] had previously published results from comparing GDX and AA in OpEx, at ICAART-2011; and the first results from AA in human-agent studies were then published in a 2011 IJCAI paper [9], in which AA was demonstrated to dominate not only humans but also GDX and ZIP. For consistency with what was by then a well-established methodology, in De Luca's experiments the $P_0$ value was static for sustained periods with occasional "shock" step-changes to different values. Continuing the tradition established by the IBM authors, the abstract of [9] claimed supremacy for AA: "*We… demonstrate that AA's performance against human traders is superior to that of ZIP, GD, and GDX. We therefore claim that… AA may offer the best performance of any published bidding strategy*". And, until the publication of Vach's 2015 MSc thesis [25], that claim appeared to be plausibly true.

## 2.4 Actually, AA doesn't dominate: Vach, 2015; Cliff, 2019.

Vach's Master's Thesis [25] tells the story of his design of a new trading strategy based on ZIP and called ZIPOJA, which he then tested against AA, GDX, and ZIP. The testing revealed that ZIPOJA did not consistently outperform any of the three pre-existing strategies. But, in the course of that testing, as Vach checked and calibrated his implementations of the three pre-existing strategies, he found that AA could fail to dominate ZIP or GDX, depending on the proportions of the two strategies in the market: this runs counter to the established story that AA is the best-performing strategy. Tables 6.2 and 6.3 on p.47 of Vach's thesis show results from tests in which the performance of two trading strategies were tested in trials with proportions of the two trader strategies set at 6:0, 5:1, 4:2, 3:3, 2:4, 5:1, and 0:6. The ratios 6:0 and 0:6 are homogenously populated by one strategy or the other and hence are of little interest, because that single strategy necessarily dominates in those markets. In Vach's Table 6.2, AA is outperformed by ZIP when the ZIP:AA ratio is 1:5 – i.e, if one in six of the traders in the market are ZIP with the rest AA, then the ZIP traders will outperform the AAs: the efficiency of the ZIP traders was 99.5% while the efficiency of the AAs was 88.5%. In Vach's Table 6.3, AA is outperformed by GDX when the GDX:AA ratio is 3:3, 2:4, and 1:5.

Vach then performed three-way simulations systematically varying the ratios of AA:GDX:ZIP over all possible permutations and, in his Fig.6.1i [25, p.53] he shows a 2D simplex diagram which summarizes those results: a 28-node regular isometric mesh is drawn over the surface of the simplex as a co-ordinate frame, and AA is the dominant strategy in only 11 of those 28 nodes. Each of the three strategies is by definition dominant at the node representing a homogeneous ratio (i.e., either 1:0:0 or 0:1:0 or 0:0:1), so AA actually only dominates at 10 of the 25 nodes where it is actually contesting with the other two strategies: ZIP dominates one of the remaining nodes; and GDX dominates the remaining 14.

In a final four-way study, with AA, GDX, ZIP, and ZIPOJA competing against each other, Vach [25, Table 6.7, p.60] declares GDX the overall winner although in that experiment the scores of GDX and AA are sufficiently close that, in our opinion, the difference between the two may not be statistically significant. Nevertheless, it is undeniable that in Vach's four-way study AA again fails to clearly dominate. To the best of our knowledge, Vach's results are the first such exhaustive study of AA's performance as the number and proportion of competitor strategies is systematically varied, and he was the first to demonstrate that AA is in fact not the best-performing strategy.

Subsequently Cliff [6] set out to replicate and extend Vach's results, using a finer-grained analysis, varying the proportions of AA, SHVR, ZIP, and ZIC, and also studying the effects of altering other aspects of the experiment design such as whether the replenishment of assignments to the traders is periodic or continuous-stochastic (as in [4]); and whether the equilibrium price $P_0$ is largely constant with occasional shock-jumps, or continuously varying according to price-movements taken from real-world markets. Cliff's results from conventional S'62-style experiments, with periodic replenishment and with $P_0$ largely constant, confirmed the established view: when AA was tested in the kind of simple market environment as has traditionally been used in the previous literature, AA scored just as well as well-known other trading strategies and was not dominated by them.

But, merely by altering the nature of the market environment to have continuous stochastic replenishment (which is surely what happens in real markets) and to have the equilibrium price $P_0$ continuously varying over time (which is also surely what happens in real markets), Cliff's results from AA became very poor indeed. Cliff [6] wrote:

> "It seems very hard to avoid the conclusion that AA's success as reported in previous papers is largely due to the extent to which its internal mechanisms are designed to fit exactly the kind of experiment settings first introduced by Vernon Smith: AA is very well suited to situations in which all assignments are issued to all traders simultaneously, and in which the equilibrium price remains constant for sustained periods of time, with only occasional step-change "shocks". Real markets are not like this, and

> when AA is deployed in the more realistic market setting provided by BSE, its dominance disappears."

Cliff did not test AA against GDX, but we do here. The results that we present in Section 4 demonstrate that actually, even in the S'62 style of experiment that AA was first tested in, if it actually is tested exhaustively across a wide range of proportions, then AA can be outperformed by trading algorithms that predated it, specifically by GDX. Before that, in Section 3 we briefly discuss the issue of modifying AA to operate in realistic LOB-based markets.

## 3 Modifying AA for LOB Markets

Taking the AA algorithm and attempting to run it in a LOB-based market reveals the extent to which AA seems designed to fit very well in the Smith'62 style of experiments with periodic replenishment, and is less well suited to a continuously varying market dynamic. In brief, AA's internal mechanisms revolve around three questions that each AA trader attempts to answer: (1) What is my best estimate of the current equilibrium price $P_0$? (2) What is my best estimate of the current volatility of transaction prices around $P_0$? And (3) is the limit price on my current assignment intramarginal (i.e., could be sold/bought at $P_0$ and still make a profit) or extramarginal? For its estimate of $P_0$, the original AA trader computes a moving average of recent transaction prices. For its volatility estimate, it computes Smith's $\alpha$ metric, taking the difference between recent transaction prices and the trader's current estimate of $P_0$ (i.e., ignoring any trend in $P_0$, which is safe to do if, as in the S'62 experiments, $P_0$ changes rarely or never). Deciding on whether the current assignment is intra/extra marginal is done by comparing its limit price to its $P_0$ estimate.

In MAA, our modified implementation of AA, these questions can instead each be answered by reference to information that is routinely available from an exchange: the LOB and the exchange's "tape" (the record of timestamped transactions). $P_0$ can be better estimated by using the volume-weighted mid-price at the top of the book (known as the *microprice*: see e.g. [2, 20]): this is a better metric because it can be sensitive to shifts in the $P_0$ value *before* any transactions go through that reflect the shift. Volatility can be estimated by reference not to only the current estimate of $P_0$ but also to BSE's tape data: a time-series of transaction-price values correlated with a time series of microprice values is better to use in situations where the $P_0$ value is continuously changing: for each transaction on the tape, the microprice at the time of that transaction (or immediately before) is the better reference value for calculating Smith's $\alpha$. Extra-/intramarginality is still decided by reference to the trader's $P_0$ estimate, but in MAA that estimate can come from the microprice.

Previous authors have also needed to adapt AA for LOB-based markets: De Luca [8-10] and Vach [25] each used AA in the *OpEx* simulator, and Stotter *et al.* [20, 21] used AA in the *ExPo* simulator. However, the modified AA proposed here is novel insofar as prior authors don't report using the exchange's tape data or the microprice.

There is a tension between modifying AA in an attempt to better fit it to a LOB-based market, and making claims about AA's poor performance in those markets: the more heavily AA is modified, the more one is open to accusations that the modifications themselves are the cause of the poor performance, rather than that poor performance being a reflection of the original AA being badly-suited to LOB markets. For that reason, in this paper, we keep AA very close to the original, using only the microprice modification in generating the results presented here.

The Python source-code used to generate the results in this paper has been made publicly available on the main BSE GitHub site [1].

## 4 Exhaustive Testing of AA

### 4.1 Market Supply and Demand Schedules

Vytelingum [26, 27] tested AA using the methods first established by Smith [18] and then followed by all of the key papers reviewed in Section 2: he did some studies with markets in which the supply and demand schedules (SDSs) were constant for the duration of each experiment, which we will refer to as *static markets*; and he did other studies in which part-way through the experiment there was a sudden "market-shock" change from the initial static SDS to some other static SDS that remained in place from the point of the shock to the end of the experiment – we will refer to those experiments as *market shocks*. Vytelingum studied AA's response in four static SDSs, which he referred to as M1, M2, M3, and M4; and his market shock studies involved switching from one of these four to one of the three other SDSs. The market shock studies were referred to using multi-character codes of the form MS*nm* where *n* is the single-digit identifier of the initial static SDS, and *m* is the single-digit identifier of the static SDS that is switched to at the time of the shock. For example, MS31 denotes an experiment in which the traders are initially given allocations according to M3, which switches to M1 at the point of the shock-change. Each of the experiments were conducted over 20 trading periods or "days", and when shocks were imposed they occurred at the start of Day 11 (i.e., halfway through the session). After carrying out preliminary tests on the SDSs used by Vytelingum, we decided that the Vytelingum's market-shock scenarios were not sufficient to completely test the algorithms: each trading algorithm adapted relatively quickly to a single shock, and hence to fully compare the trading strategies we decided to introduce more challenging markets, some containing more shocks, and also some with a continuously changing equilibrium price.

**Static Markets.** First we tested the trading agents using static SDSs based on M1 to M4 as used by Vytelingum [26, 27]: the supply and demand curves for each market are shown in Figure 4.1.

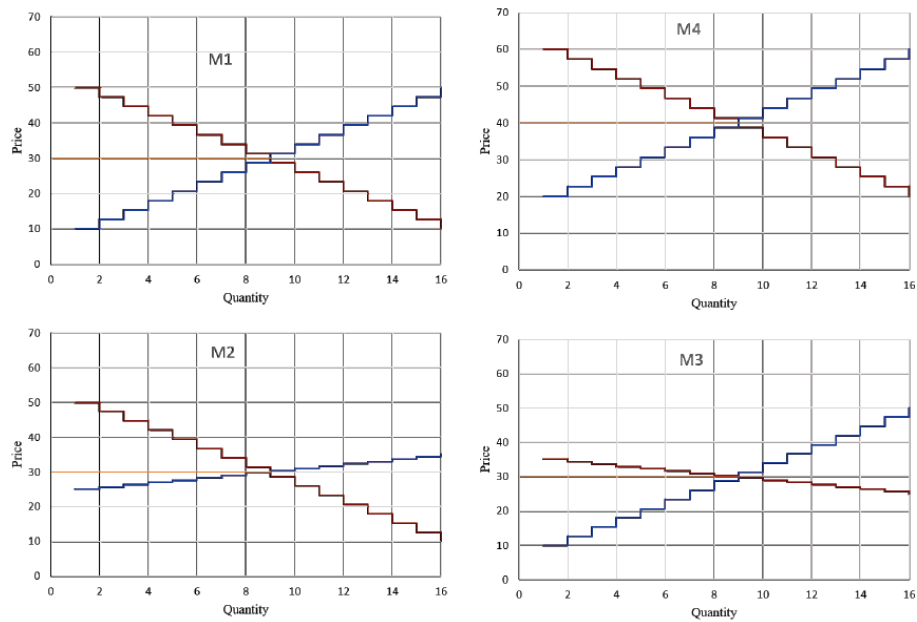

**Fig. 4.1.** Supply and Demand curves for M1, M2, M3, and M4. The expected equilibrium price is marked: for M1, M2, and M3 it is 30; for M4 it is 40.

**Complex Markets.** We tested market shocks introduced in the manner described by Vytelingum [26, 27], specifically MS14, MS21, MS31, MS23, and MS1231. We then also explored the responses of the traders in situations where all prices on assignments came from M1 with a time-varying offset function $F(t)$ added to them over the course of the experiment. We refer to these as follows:

- M6: $F(t) = c \sin(t/30)$ (a sinusoid of constant amplitude and frequency).
- M7: $F(t) = ct(1+\sin(wt))$ (a sinusoid of increasing amplitude and frequency).
- M8: $F(t) = (t\,\%75)/2$ (a sawtooth wave of constant amplitude and frequency).
- M9: $F(t) = c \, \text{sgn}(\sin(t/30))$ (a square wave of constant amplitude and frequency).

### 4.2 Verification

Although the source-code for ZIP was published as an appendix to the technical report that introduced that algorithm [3], no standard reference implementations exist for either GDX or AA: in both cases, the papers that introduced the algorithm gave verbal descriptions of how the algorithm works, along with associated equations. To verify that the implementations of the algorithms used in this paper are correct, we conducted experiments whose purpose was to replicate results achieved in the algorithms' original papers. Full details of these verification experiments are given in [19], to which the reader is referred for further details. It is sufficient to note here that our results from GDX and AA were in both cases very close but not identically equal to the results published in the relevant original paper. We believe that the differences in results are more likely to be due to differences in test environment than due to any problems with our implementation of the algorithms. The original papers for GDX and AA say very little about the nature of the market simulator that was used to generate the results. We use the public-domain BSE simulator, but GDX was tested on IBM's in-house Magenta market simulator, about which nothing was ever published; and Vytelingum [26, 27] discusses his own market simulator only in very scant terms. Thus, to the best of our ability, we believe the implementations of GDX and AA used here to be faithful to the original specification. The source-code used to generate the results in this section (which summarizes the results presented in [19]) has been made publicly available in the BSE GitHub repository [1]: see the script `snashall2019.py`.

### 4.3 Experiment Design

As each market scenario has a different expected profit, and we used allocative efficiency as the measure of performance, for ease of comparison across all experiments. As described in above, this is the percentage of the maximum expected profit the algorithm has been able to extract from the market.

Each trial involves 16 traders on each side (32 in total). It is known that the different trader ratios can have a profound and significant effect on their respective performance. For example, a single ZIP agent in a market populated by ZIC agents will do exceedingly well, however a single ZIP agent in a market otherwise saturated with GDX agents will not do as well. To eliminate this effect, the experiments here are conducted with every possible permutation of trader ratios, and the results are averaged over every experiment. We conducted 100 i.i.d trials per ratio, which equates to around 2 million trading days in total. We then compute summary statistics, such as average efficiency, across all trials, and present those in tabular form; for ease of identification we use a bold-face font to highlight the highest (best) value in each row.

### 4.4 Results

**Static markets.**
Tables 4.4a to 4.4d show various results from simple static market experiments. Tables 4.4a and 4.4b are from S'62-style experiments in which the assignments to buy and sell are refreshed periodically, with all traders receiving their updates simultaneously. Tables 4.4c and 4.4d are from experiments in which the assignments are instead stochastically drip-fed into the population of traders in a continuous-replenishment approach as described by [4].

Table 4.4a shows results from markets populated by mixes of AA, ASAD, GDX, and ZIP traders, with periodic allocation: the overall average has AA scoring a shade higher than GDX, but GDX scores slightly higher than AA in markets M1 and M3. Because ASAD and ZIP are very closely related, and AA is arguably also an extension of the basic ZIP algorithm (i.e., it shares the same heuristic decision tree, but adds sophistication in how the trader's profit margin is altered over time), it might be argued that the experiments summarized in Table 4.4a are essentially GDX *versus* three variants of ZIP-style algorithms. Indeed, if (as we believe) it is fair to characterize ASAD as ZIP with extensions to detect shock-changes in market prices and act appropriately, the absence of any market shocks in these simple experiments mean that ZIP and ASAD are essentially functionally identical. To increase the heterogeneity, we ran the same

experiments again but replaced ZIP with the simpler, more noisy, ZIC strategy: results from that are shown in Table 4.4b. Now only in M5 does AA still dominate: GDX wins in M1, M2, and M3.

| Market | AA | ASAD | GDX | ZIP |
|--------|--------|--------|--------|--------|
| M1 | 97.03 | 79.97 | **98.85** | 80.95 |
| M2 | **103.21** | 54.57 | 100.01 | 55.92 |
| M3 | 99.13 | 85.80 | **99.41** | 85.23 |
| M5 | **99.52** | 60.32 | 97.13 | 58.67 |
| Average | **99.73** | 70.16 | 98.85 | 70.19 |

**Table 4.4a.** Efficiencies in AA/ASAD/GDX/ZIP experiments with periodic replenishment.

| Market | AA | ASAD | GDX | ZIC |
|--------|--------|--------|--------|--------|
| M1 | 94.47 | 87.90 | **96.78** | 59.82 |
| M2 | 96.51 | 75.98 | **100.2** | 77.49 |
| M3 | 94.39 | 84.69 | **94.77** | 43.06 |
| M5 | **95.18** | 72.53 | 91.58 | 50.12 |
| Average | 95.14 | 80.27 | **95.82** | 57.64 |

**Table 4.4b.** Efficiencies in AA/ASAD/GDX/ZIC experiments with periodic replenishment.

When we switch from periodic to continuous replenishment, the summary data show broadly the same pattern: Table 4.4c shows that when GDX is pitted against three ZIP-style strategies, it is out-scored by AA in half of the markets studied, and AA scores best overall; but Table 4.4d shows that when we replace ZIP with ZIC, this alters the market dynamics and GDX now dominates in three of the four markets and also in aggregate score.

| Market | AA | ASAD | GDX | ZIP |
|--------|--------|--------|--------|--------|
| M1 | 92.70 | 80.34 | **98.43** | 80.26 |
| M2 | **105.27** | 60.71 | 105.24 | 60.53 |
| M3 | 100.66 | 87.17 | **103.80** | 87.43 |
| M5 | **97.40** | 52.54 | 92.01 | 52.53 |
| Average | 99.01 | 70.19 | **99.87** | 70.19 |

**Table 4.4c.** Efficiencies in AA/ASAD/GDX/ZIP experiments with continuous replenishment.

| Market | AA | ASAD | GDX | ZIC |
|--------|--------|--------|--------|--------|
| M1 | 88.58 | 89.33 | **94.60** | 59.82 |
| M2 | 99.71 | 82.41 | **102.07** | 77.49 |
| M3 | 97.87 | 97.99 | **102.91** | 43.05 |
| M5 | **94.94** | 83.44 | 93.18 | 50.18 |
| Average | 95.28 | 88.29 | **98.19** | 57.64 |

**Table 4.4d.** Efficiencies in AA/ASAD/GDX/ZIC experiments with continuous replenishment.

**Complex Markets.** Tables 4.4e and 4.4f respectively summarize our results from testing in the complex markets introduced in Section 4.1, with ZIP and ZIC. Here there are no subtleties in the outcomes: GDX is clearly dominant in all markets reported on in Table 4.4e, and again in Table 4.4f.

The nonparametric Wilcoxon-Mann-Whitney U-Test was used to evaluate the statistical significance of the differences in scores between AA and GDX in Tables 4.4e and 4.4f. This indicated that the difference is significant in all cases except for M2 in Table 4.4e, and for M7 in Table 4.4f. In all cases where a significant difference was detected, GDX had the better score: for further details see [19, pp.29-30].

| Market | AA | ASAD | GDX | ZIP |
|--------|------|------|------|------|
| MS14 | 93.40 | 75.36 | **96.22** | 73.30 |
| MS21 | 91.18 | 73.17 | **94.43** | 73.24 |
| MS31 | 93.06 | 83.55 | **98.10** | 83.63 |
| MS23 | 104.46 | 50.72 | **105.79** | 50.69 |
| MS1231 | 102.18 | 87.76 | **104.37** | 84.51 |
| M6 Sin | 70.53 | 55.62 | **72.87** | 55.56 |
| M7 | 99.21 | 92.09 | **102.18** | 91.12 |
| M8 Saw | 86.52 | 91.79 | **95.80** | 91.71 |
| M9 Sqr | 69.30 | 63.46 | **74.61** | 63.70 |
| Average | 89.98 | 74.84 | **93.82** | 75.92 |

**Table 4.4e.** Efficiencies in complex markets, with ZIP.

| Market | AA | ASAD | GDX | ZIC |
|--------|------|------|------|------|
| MS14 | 85.08 | 87.14 | **91.12** | 42.05 |
| MS21 | 92.68 | 89.04 | **96.76** | 68.45 |
| MS31 | 89.64 | 91.16 | **95.83** | 51.50 |
| MS23 | 100.62 | 97.74 | **104.57** | 61.76 |
| MS1231 | 94.34 | 95.52 | **101.52** | 60.56 |
| M6 Sin | 69.93 | 65.98 | **72.89** | 44.58 |
| M7 | 75.86 | 69.58 | **77.71** | 50.68 |
| M8 Saw | 86.77 | 91.59 | **95.02** | 64.69 |
| M9 Sqr | 70.93 | 64.35 | **72.56** | 53.03 |
| Average | 85.10 | 83.57 | **89.78** | 55.25 |

**Table 4.4f.** Efficiencies in complex markets, with ZIC.

To summarize, the results presented here show that GDX routinely and reliably dominates AA. That reinforces the message from Vach [25] and Cliff [6]: AA does not dominate.

# 5 METHODOLOGICAL ISSUES

Having demonstrated that AA does not always dominate other trading strategies, it is worth reflecting on the methods that have been used here, how they compare to current real-world financial markets, and how they compare to the S'62 methods that were described in Section 2.

### 5.1 Real-World Relevance

BSE, the open-source public-domain CDA market simulator that we have used as the platform for our studies, was introduced in Section 2.2. There are numerous differences between BSE and real financial markets: BSE is not intended to be a perfect imitation of a real stock exchange; indeed it was initially created to support graduate-level teaching, conducting experiments in the same vein as S'62. It is designed to provide an environment in which experiments can be reliably repeated and controlled, rather than providing an environment which is as close as possible to real-world market scenarios.

BSE does not simulate communications latency: it assumes all traders receive information updates from the exchange instantaneously, and similarly it assumes that any message sent by a trader to the exchange takes zero time to arrive. In the real world, things are not so simple: it takes finite time for the market information published by an exchange to reach any given trader, and it takes finite time for a trader's order to reach the exchange. Communications latency of this form can play a large part in the performance of an algorithm. For example, if a trader is designed to execute an arbitrage strategy (that is to take advantage of price difference between markets, e.g. buying something on Exchange A and then immediately selling it on Exchange B for a higher price), the trader may have only milliseconds to act before parity is restored.

Another form of real-world latency that BSE fails to simulate is the processing latencies of the trading algorithms themselves, i.e. their *reaction-time*. The reaction time of an algorithm can play just as an important role in its performance as the communications delay. In real markets, the traders must respond as quickly as possible to the market. If, for example, ZIP is able to respond more quickly than GDX and therefore put an ask/bid in earlier, it will steal the opportunity for GDX to make a trade. In the currently available version of BSE, this is not captured, because each algorithm is allowed to take as long as it wants to respond (the simulation is single-threaded, and simulates parallel activity by allowing all traders to settle on a response to any change in the market before processing the responses of each trader). Each algorithm implemented in BSE was written with the assumption that the state of the market will not change while the algorithm is 'thinking'. In the real world this is absolutely not the case. The market is changing constantly, and any trader can submit a fresh ask or bid at any time. In the case of both GDX and AA, their designs mean that each time a new ask/bid is submitted, they must start their processing again from the beginning. GDX must re-compute its belief function, and AA must re-compute all of its various calculations. Due to the frequency of submission of quotes in CDAs dominated by 'high frequency' traders, it could be argued that neither AA nor GDX would ever be quick enough to submit an ask/bid before the market has significantly changed again, forcing them to re-start their calculations. This lack of any modelling of reaction-times runs the risk of incorrect conclusions about dominance relationships being drawn when trading agents are evaluated only in the simple S'62 style scenarios used here.

It could be argued that the aim of algorithms such as AA is not necessarily to perform well in the real world, but instead just to beat their competitor algorithms in the kind of comparative studies described here (on in actual international trading-agent contests, popular with academics around the world, as described in e.g. [28]). We don't agree with that view: our opinion is that if a trading agent does well in academic research contests but is not applicable in real-world deployments, it is of little interest to us.

To test the extent to which actual reaction-times could affect our results, we conducted an experiment where, for the various trading strategies used here, we measured how long it takes for our implementation of that strategy in BSE to respond with an order after it is sent updated market information. We conducted this in four of the markets used previously, over 500 trials each, and with a fixed ratio of 5 buyers and 5 sellers using each strategy: the results are shown in Table 5.1.

| Market | AA | ASAD | GDX | ZIP |
|--------|-----|------|-----|-----|
| M7 czy | $6.20\mu s$ | $5.00\mu s$ | $80.83\mu s$ | $5.27\mu s$ |
| M6 sin | $5.14\mu s$ | $5.69\mu s$ | $87.56\mu s$ | $4.73\mu s$ |
| M1 | $6.27\mu s$ | $5.33\mu s$ | $57.18\mu s$ | $4.94\mu s$ |
| MS23 | $5.96\mu s$ | $5.44\mu s$ | $87.43\mu s$ | $5.44\mu s$ |
| Average | $5.89\mu s$ | $5.37\mu s$ | $78.25\mu s$ | $4.96\mu s$ |

**Table 5.1a.** Efficiencies in complex markets, with ZIP.

Clearly, the implementation of GDX used here is consistently an order of magnitude slower than the other strategies when deciding on its next ask/bid price. In the experiments in this paper, of course this is of no significance. However, in a real market, this could very easily cause the GDX strategy to fail to generate a profit because AA, ZIP, and ASAD all have at least 10 opportunities to trade while GDX is calculating a single ask/bid price. BSE does not yet include functionality for multi-threading which makes it a poor platform for studying time-sensitive responses. We should also note here that in our implementation of these algorithms very little thought was given to run-time efficiency. GDX requires the creation of a 2D data structure containing expected values, and also computes the argmax of a relatively complicated function, which is understandably slow as the program must traverse every possible input. Our implementation of this is simplistic and written to be easy to follow, rather than to be quick. A more time-efficient implementation might reduce the disparity between GDX and its competitor strategies, e.g. by using precomputed look-up-tables.

## 5.2 Perpetuation of Smith'62-Style Norms

We find it hard to avoid the conclusion that AA's success as reported in previous papers is largely due to the extent to which its internal mechanisms are designed to fit exactly the kind of experiment settings first introduced by Vernon Smith: AA is very well suited to situations in which all assignments are issued to all traders simultaneously, and in which the theoretical equilibrium price remains constant for sustained periods of time, with only occasional step-change "shocks". Real markets are not like this, and as Cliff showed in [6], when AA is deployed in the more realistic market setting provided by BSE, its dominance disappears. The novel aspect of the results we present here is that we have now demonstrated that even in the simple style of experiments that AA was first tested in, AA can be shown not to dominate if sufficiently many tests are run.

Surely then the broader methodological lesson here is that we should not allow ourselves to be seduced by results from small-scale studies in minimally simple approximations to real-world markets. Smith developed his experimental methods in the late 1950's when there were no realistic alternative ways of doing things. Running experiments with human subjects is laborious and slow, but experiments in electronic markets populated entirely by robot traders can proceed in appropriate simulators at speeds much faster than real-time, and are "embarrassingly parallelizable": the more computational cores or virtual machines dedicated to the task, the faster the exhaustive experiments complete.

At this point in time, 20% of our way into the 21$^{st}$ Century, surely trading-agent researchers should collectively abandon the simple minimal test-environments that worked well for Vernon Smith in the middle of the 20$^{th}$ Century and instead start to tolerate the minor inconvenience of running very large numbers of trials on reasonably accurate simulations of realistic market situations: the methods used here should be the norm, not the exception. The availability of open-source public-domain exchange simulators such as BSE as a common platform for experiments and as a source of reference implementations, coupled with readily available cheap cloud-computing for doing the necessary processing, means that there are now really no excuses for not doing so.

## 6 CONCLUSIONS

The design of trading agents has been a research topic within AI/ML for over two decades, with the initial work taking place in the research labs of major technology companies such as IBM and HP, and at peak involved 20 or more teams of researchers around the world, some of whom would compete in the various trading agent competitions (TAC) held at AAAI and AAMAS conferences (see e.g. [28] for a summary of TAC research). Anyone reading the published literature might reasonably come to the conclusion that Vytelingum's AA strategy [26,27] has remained unchallenged for more than a decade as the best-known public domain strategy for trading in continuous double auctions (CDAs) such as those found in the global financial markets; and in that sense CDA trading-agent design may have been thought by many to have been consigned to AI's list of "solved problems".

In this paper we have demonstrated that the apparent success of AA was in fact due to it not having been tested sufficiently. Our experiments were inspired by, and extend, those of Vach [25] and Cliff [6] but the AA source-code we used to generate the results presented here was developed independently of those two authors' work. That is, there are now three independent studies that each indicate AA to not be a dominant strategy.

We do not intend this paper to cast any doubts on the scientific or engineering merits of the previous work that we here call into question. In the decade that has passed since Vytelingum first published his AA work, the continuing Moore's Law fall in the real cost of computing hardware, combined with the rise of cheap and readily scalable remotely accessed cloud computing, gives today's researchers access to compute-power that would arguably have been unimaginable, or at least prohibitively expensive, over a decade ago when the first tests were being run on AA. As our brute-force exhaustive evaluation of AA competing with other strategies across all possible permutations shows, we are now in the lucky position to be able to ask, and to answer, questions that would not have been practicable to attempt to explore 10 or 15 years ago.

And the conclusion that we have arrived at is this: AA is clearly not the dominant, best-performing CDA trading strategy; in the experiments reported here, it is outperformed by GDX (as in [25]), and in [6] it is outperformed by ZIP. This reverses the solidly-stated conclusions of previous papers, asserting AA's dominance.[2]

Methodologically, all of the studies reviewed here (including our own experiments) are firmly in the same minimally simple frame of reference first established by Vernon Smith in his 1962 experiments: agents are assigned a right to buy or sell only a small number of items (typically only one) at any one time; and none of AA or ZIP or ASAD or GDX or ZIC have any sense of size-sensitivity (larger-sized orders being more significant than smaller ones) nor of time-sensitivity (some orders being more urgent to get executed than others). The strategies that have been studied in the CDA trading-agent literature are (with the notable exceptions of the famous *Kaplan Sniper* algorithm described in [17]; and Gjerstad's *HBL* strategy [12]) almost exclusively focused solely on price. Yet traders in real-world markets need to reason about price, and quantity, and time, making dynamic tradeoffs as the market moves over time. There is a clear need for further research directed at creating such more sophisticated, and hence more real-world-relevant, trading strategies, and then comparing and evaluating them appropriately.

But, as we have argued here, there is also a clear need for future research to be conducted in such a way that erroneous conclusions are less likely to be drawn and promulgated. One way of doing that is to burn through very large numbers of compute-cycles, working exhaustively through all permutations of different strategies that might reasonably be found in a CDA market somewhere sometime. A CDA trading strategy should only be described as dominant, or the best-performing, if it really is; and sometimes, more often than not, the only way of determining that is to run an awful lot of experiments. If all those experiments take a lot of money to run on a lot of machines, we just need to bear that cost; and if they take a long time to run, we just need to be patient. But, thankfully, the availability of low-cost cloud computing services means that we don't need to spend as much money on supercomputers, and nor do we need to wait as long as if we only had a few cores available. Now that the results we've presented here have overturned long-held beliefs about which is the best-performing public-domain trading strategy, running large-scale exhaustive experiments on contemporary scalable cloud services (or equivalent locally-available hardware) seems like the only reasonable way forward in future.

---

[2] At least two of those papers were co-authored by one of us, Dave Cliff. So this present paper is offered as something of a *mea culpa* from Cliff.